\documentclass[prd,superscriptaddress,amsfonts,amssymb,amsmath,showpacs,twocolumn]{revtex4-2}
\usepackage{bm}
\usepackage{amsfonts}
\usepackage{latexsym}
\usepackage[latin1]{inputenc}
\usepackage{graphicx}
\usepackage{amsmath}
\usepackage{palatino}
\usepackage{ragged2e}
\usepackage{mathpazo}
\usepackage{textcomp}
\usepackage{float}
\usepackage{booktabs}
\usepackage{dcolumn}
\usepackage{multirow}
\usepackage{hyperref}
\hypersetup{colorlinks,citecolor=blue}
\usepackage{amsmath}
\usepackage{xcolor}
\usepackage{orcidlink}
\usepackage[caption=false]{subfig}
\usepackage{commath}
\def\jnl@style{\it}
\def\aaref@jnl#1{{\jnl@style#1}}

\def\aaref@jnl#1{{\jnl@style#1}}

\def\aj{\aaref@jnl{AJ}}                   
\def\apj{\aaref@jnl{ApJ}}                 
\def\apjl{\aaref@jnl{ApJ}}                
\def\apjs{\aaref@jnl{ApJS}}               
\def\apss{\aaref@jnl{Ap\&SS}}             
\def\aap{\aaref@jnl{A\&A}}                
\def\aapr{\aaref@jnl{A\&A~Rev.}}          
\def\aaps{\aaref@jnl{A\&AS}}              
\def\mnras{\aaref@jnl{Mon.~Not.~Roy.~Astron.~Soc.}}             
\def\prd{\aaref@jnl{Phys.~Rev.~D}}        
\def\prc{\aaref@jnl{Phys.~Rev.~C}}  
\def\prl{\aaref@jnl{Phys.~Rev.~Lett.}}    
\def\qjras{\aaref@jnl{QJRAS}}             
\def\skytel{\aaref@jnl{S\&T}}             
\def\ssr{\aaref@jnl{Space~Sci.~Rev.}}     
\def\zap{\aaref@jnl{ZAp}}                 
\def\nat{\aaref@jnl{Nature}}              
\def\aplett{\aaref@jnl{Astrophys.~Lett.}} 
\def\apspr{\aaref@jnl{Astrophys.~Space~Phys.~Res.}} 
\def\physrep{\aaref@jnl{Phys.~Rep.}}      
\def\physscr{\aaref@jnl{Phys.~Scr}}       
\def\commat{\aaref@jnl{Comm.~Math.~Phys.}}              
\def\science{\aaref@jnl{Science}}               
\def\cqg{\aaref@jnl{Classical Quant.~Grav.}}            
\def\jpcs{\aaref@jnl{JPCS}}                                     
\def\ijmpd{\aaref@jnl{Int.~J.~Mod.~Phys.~D}}                    
\def\grg{\aaref@jnl{Gen.~Relat.~Gravit.}}               
\def\rpp{\aaref@jnl{Rep.~Prog.~Phys.}}          
\def\npa{\aaref@jnl{Nucl.~Phys.~A}}        
\def\lrr{\aaref@jnl{Living Rev.~Rel.}}                   
\def\jcap{\aaref@jnl{J.~Cosmology Astropart.~Phys.}}    
\def\rmp{\aaref@jnl{Rev.~Mod.~Phys.}}   
\def\epjc{\aaref@jnl{Eur.~Phys.~J.~C}}

\allowdisplaybreaks[1]

\begin{document}

\color{black}       

\title{Constrained evolution of effective equation of state parameter in non-linear $f(R, L_m)$ dark energy model: Insights from Bayesian analysis of cosmic chronometers and Pantheon samples}

\author{N. Myrzakulov\orcidlink{0000-0001-8691-9939}}
\email[Email: ]{nmyrzakulov@gmail.com}
\affiliation{L. N. Gumilyov Eurasian National University, Astana 010008,
Kazakhstan.}
\affiliation{Ratbay Myrzakulov Eurasian International Centre for Theoretical
Physics, Astana 010009, Kazakhstan.}

\author{M. Koussour\orcidlink{0000-0002-4188-0572}}
\email[Email: ]{pr.mouhssine@gmail.com}
\affiliation{Quantum Physics and Magnetism Team, LPMC, Faculty of Science Ben
M'sik,\\
Casablanca Hassan II University,
Morocco.}

\author{Alnadhief H. A. Alfedeel\orcidlink{0000-0002-8036-268X}}%
\email[Email: ]{aaalnadhief@imamu.edu.sa}
\affiliation{Department of Mathematics and Statistics, Imam Mohammad Ibn Saud Islamic University (IMSIU),\\
Riyadh 13318, Saudi Arabia.}
\affiliation{Department of Physics, Faculty of Science, University of Khartoum, P.O. Box 321, Khartoum 11115, Sudan.}
\affiliation{Centre for Space Research, North-West University, 2520 Potchefstroom, South Africa.}

\author{Amare Abebe}
\email[Email: ]{amare.abebe@nithecs.ac.za}
\affiliation{Centre for Space Research, North-West University, 2520 Potchefstroom, South Africa.}
\affiliation{National Institute for Theoretical and Computational Sciences (NITheCS), 3201 Stellenbosch, South Africa}

\date{\today}
\begin{abstract}

We conduct a Bayesian analysis of recent observational datasets, specifically the Cosmic Chronometers (CC) dataset and Pantheon samples, to investigate the evolution of the EoS parameter in dark energy models. Our study focused on the effective EoS parameter, which is described by the parametric form $\omega_{eff}=-\frac{1}{1+m(1+z)^n}$, where $m$ and $n$ are model parameters. This parametric form is applicable within the framework of $f(R,L_m)$ gravity, where $R$ represents the Ricci scalar and $L_m$ is the matter Lagrangian. Here, we examine a non-linear $f(R,L_m)$ model characterized by the functional form $f(R,L_m)=\frac{R}{2}+L_m^\alpha$, where $\alpha$ is the free parameter of the model. We examine the evolution of several cosmological parameters, including the effective EoS parameter $\omega_{eff}$, the deceleration parameter $q$, the density parameter $\rho$, the pressure $p$, and the statefinder parameters. Our analysis revealed that the constrained current value of the effective EoS parameter, $\omega_{eff}^{0}=-0.68\pm0.06$ for both the CC and Pantheon datasets, points towards a quintessence phase. Moreover, at redshift $z=0$, the deceleration parameter, $q_0 = -0.61^{+0.01}_{-0.01}$, indicates that the present Universe is undergoing accelerated expansion. 


\end{abstract}

\maketitle

\section{Introduction}\label{sec1}

Recent astrophysical observations and data have provided compelling evidence for the expanding Universe. The phenomenon of cosmic acceleration is supported by various sources such as high redshift Supernovae (SNe), Cosmic Microwave Background Radiation (CMBR), Wilkinson Microwave Anisotropy Probe (WMAP), Baryonic Acoustic Oscillations (BAOs), and Large scale Structure (LSS) \cite{Riess,Perlmutter,C.L.,R.R.,D.N.,D.J.,W.J.,T.Koivisto,S.F.}. These observations have revealed the presence of a mysterious component known as Dark Energy (DE), which permeates the Universe and contributes to approximately 70\% of its total energy budget. DE possesses the unique property of exerting a strong negative pressure, making it distinct from conventional forms of matter and energy. Its enigmatic nature adds to the fascination and complexity of understanding the fundamental workings of our Universe. DE plays a crucial role in the accelerated expansion of the Universe and can be characterized by its Equation of State (EoS) parameter $\omega = p/\rho$, where $p$ represents the pressure and $\rho$ represents the energy density. Numerous studies have demonstrated that the Universe undergoes accelerated expansion when the EoS parameter approaches $\omega=-1$ \cite{S.W.,E.J.}. In certain scenarios, DE exhibits phantom-like behavior, indicated by $\omega<-1$. A Universe governed by phantom DE is predicted to experience a future singularity called cosmic doomsday or the big rip, where the Universe is torn apart \cite{Phan1,Phan2,Phan3}. To understand and explain the nature of DE and its connection to late-time acceleration, researchers have explored modified theories of gravity. These theories offer an intriguing alternative to conventional approaches and hold promise in addressing the puzzles of cosmic acceleration and quintessence \cite{Quin}.

In the past few decades, numerous generalizations of Einstein's field equations have emerged, each offering unique perspectives on the nature of gravity. One such theory is the $f(Q)$ theory of gravity (where $Q$ is the non-metricity scalar) \cite{Q0,Q1}, which has gained attention in recent years. This theory represents a generalized version of symmetric teleparallel gravity, where the conventional Levi-Civita connection is replaced with the Weyl connection. What makes this theory particularly intriguing is its potential to explain the current cosmic acceleration without the need for DE. As a result, a significant amount of research has been conducted by various scholars in this field \cite{Q2,Q3,Q4,Q5,Q6,Q7,Q8}, furthering our understanding of the $f(Q)$ theory of gravity and its implications for the dynamics of the Universe. In recent years, researchers have turned their attention to another extended theory of gravity called the $f(R)$ theory \cite{H.A.,R.K.,H.K.}. This theory expands upon the standard Einstein-Hilbert action by introducing a function of the Ricci scalar $R$. By incorporating the term $1/R$, which becomes significant at small curvatures, the $f(R)$ theory offers a potential explanation for cosmic acceleration. The cosmological relevance of $f(R)$ models has made the $f(R)$ theory particularly appealing in understanding the late-time expansion scenario \cite{Carr, Cap, LAM}. Furthermore, viable $f(R)$ gravity models have been explored within the framework of solar system tests \cite{Noj, V.F., P.J., L.A.}. The observational signatures of $f(R)$ DE models, as well as the constraints imposed by the solar system and equivalence principle, have been extensively investigated \cite{Shin, Liu, Sal, Sean, Alex}. Additional $f(R)$ models have been proposed to unify early inflation with DE and to satisfy local tests \cite{Noj-2, Noj-3, G.C.}. For further insights into the cosmological implications of $f(R)$ gravity models, one can refer to the references \cite{JS, SC, RC}. The ongoing research in this field sheds light on the potential of $f(R)$ theory in addressing various cosmological phenomena.

A recent advancement in the field of gravity theories is the proposal by Harko et al. \cite{RL} of a new generalization called $f(R,L_m)$ theory of gravity. In this theory, $R$ represents the scalar curvature, while $L_m$ corresponds to the matter Lagrangian density. This extension introduces a novel approach to understanding gravitational dynamics by incorporating both the geometric property of curvature and the energy distribution described by the matter Lagrangian density. The intriguing connection between matter and geometry gives rise to an additional force perpendicular to the four-velocity vector, resulting in the non-geodesic motion of massive particles. Building upon this concept, the study of arbitrary couplings in both matter and geometry has been extended \cite{THK}. Extensive investigations into the cosmological and astrophysical implications of these non-minimal matter-geometry couplings have been carried out \cite{THK-2,THK-3,SNN,V.F.-2,V.F.-3}. It should be noted that $f(R,L_m)$ gravity models exhibit an explicit violation of the equivalence principle, which is rigorously constrained by solar system tests \cite{FR,JP}. Recently, Wang and Liao investigated the energy conditions within the framework of $f(R,L_m)$ gravity \cite{WG}. Furthermore, Goncalves and Moraes analyzed cosmological aspects by considering the non-minimal matter-geometry coupling in $f(R,L_m)$ gravity \cite{GM}. Solanki et al. \cite{Solanki} made a significant contribution to the $f(R, L_m)$ background by investigating cosmic acceleration within the framework of an anisotropic space-time with bulk viscosity. In addition, Jaybhaye et al. \cite{Jaybhaye} conducted an insightful study focused on constraining the EoS for viscous DE in the context of $f(R, L_m)$ gravity. Their research provides valuable insights into understanding the nature of DE within this specific gravitational framework, contributing to our broader comprehension of cosmic acceleration and its underlying mechanisms. These studies shed light on the intriguing implications and consequences of the interplay between matter and geometry in the context of $f(R,L_m)$ gravity. Currently, there is a growing body of literature exploring the intriguing cosmological implications of $f(R,L_m)$ gravity theory. Several studies have emerged that delve into the various aspects and consequences of this theory \cite{RL1,RL2,RL3,RL4}.

In this study, we employ the reconstruction of an effective EoS parameter to gain insights into late-time cosmic acceleration within the framework of $f(R, L_m)$ gravity. This effective EoS is not influenced by the distinct properties of individual matter field components. Instead, it is determined by model parameters that we constrain based on observational data, specifically targeting the current value of the effective EoS parameter.  Observational cosmology, a discipline dedicated to investigating the universe's structure, existence, and evolutionary processes through empirical observations, has provided a wealth of data supporting the notion of cosmic acceleration. Notably, datasets such as CMBR, Type Ia SNe, BAOs, Planck data \cite{planck_collaboration/2020}, and others have offered robust evidence for this phenomenon. To constrain our model parameters, we rely on two key datasets: the cosmic chronometers (CC) dataset and the Pantheon dataset. The CC dataset comprises 31 data points obtained from the differential age method, spanning the redshift range $0.07 < z < 2.42$ \cite{Moresco/2015,Moresco/2018}. In addition, we incorporate the recently introduced SNe Pantheon sample, consisting of 1048 data points covering the redshift interval $0.01 < z < 2.26$ \cite{Scolnic/2018}. To perform parameter estimation, we employ the MCMC ensemble sampler, as provided by the emcee library \cite{Mackey/2013}.

The structure of this paper is as follows:  In Sec. \ref{sec2}, we introduce the formalism of $f(R,L_m)$ gravity. Sec. \ref{sec3} focuses on a cosmological $f(R,L_m)$ model, where we derive expressions for the Hubble parameter and the deceleration parameter using a parametric form of the effective EoS parameter. Next, in Sec. \ref{sec4}, we determine the optimal values for the model parameters by using the observational datasets CC and Pantheon. We also examine the behavior of cosmological parameters for these constrained model parameter values. In addition, in Sec. \ref{sec5}, we investigate the statefinder parameters to discern our cosmological model from other DE models. Finally, in Sec. \ref{sec6}, we discuss our findings and provide concluding remarks. 
 
\section{The Theory of Gravity with $f(R,L_m)$ Formalism}\label{sec2}

The $f(R,L_m)$ gravity theory, originally proposed as an expansion of $f(R)$ theories, introduces a novel approach by incorporating both the Ricci scalar $R$ and the matter Lagrangian term $L_m$ as general functions within the gravitational part of the model action \cite{RL},
\begin{equation}\label{1}
S= \int{f(R,L_m)\sqrt{-g}d^4x}, 
\end{equation}
where $g$ represents the determinant of the metric, while $f(R,L_m)$ refers to an arbitrary function that incorporates both $R$ and $L_m$ as its variables. Furthermore, in this article, we will adopt natural units as our choice of measurement.

The Ricci scalar $R$ can be derived by performing a contraction of the Ricci tensor $R_{\mu\nu}$ using the metric tensor $g^{\mu\nu}$. This contraction yields the following expression for $R$,
\begin{equation}\label{2}
R= g^{\mu\nu} R_{\mu\nu},
\end{equation} 
where the definition of the Ricci tensor is given by
\begin{equation}\label{3}
R_{\mu\nu}= \partial_\lambda \Gamma^\lambda_{\mu\nu} - \partial_\mu \Gamma^\lambda_{\lambda\nu} + \Gamma^\lambda_{\mu\nu} \Gamma^\sigma_{\sigma\lambda} - \Gamma^\lambda_{\nu\sigma} \Gamma^\sigma_{\mu\lambda},
\end{equation}
where the components $\Gamma^\alpha_{\beta\gamma}$ refer to the well-known Levi-Civita connection.

The field equation for the metric tensor $g_{\mu\nu}$ can be obtained by varying the action \eqref{1},
\begin{equation}\label{5}
f_R R_{\mu\nu} + (g_{\mu\nu} \square - \nabla_\mu \nabla_\nu)f_R - \frac{1}{2} (f-f_{L_m}L_m)g_{\mu\nu} = \frac{1}{2} f_{L_m} T_{\mu\nu}.
\end{equation}

In this context, the notations $f_R \equiv \frac{\partial f}{\partial R}$ and $f_{L_m} \equiv \frac{\partial f}{\partial L_m}$ are introduced. In addition, $T_{\mu\nu}$ denotes the energy-momentum tensor for a perfect type fluid, which is defined by
\begin{equation}\label{6}
T_{\mu\nu} = \frac{-2}{\sqrt{-g}} \frac{\delta(\sqrt{-g}L_m)}{\delta g^{\mu\nu}}.
\end{equation}

The relation between the trace of the energy-momentum tensor $T$, the Ricci scalar $R$, and the Lagrangian density of matter $L_m$ can be derived by contracting the field equation \eqref{5},
\begin{equation}\label{7}
R f_R + 3\square f_R - 2(f-f_{L_m}L_m) = \frac{1}{2} f_{L_m} T,
\end{equation}
where $\square F$ represents the d'Alembertian of a scalar function $F$, given by $\square F = \frac{1}{\sqrt{-g}} \partial_\alpha (\sqrt{-g} g^{\alpha\beta} \partial_\beta F)$.

Furthermore, by taking the covariant derivative of Eq. \eqref{5}, we obtain the following result,
\begin{equation}\label{8}
\nabla^\mu T_{\mu\nu} = 2\nabla^\mu ln(f_{L_m}) \frac{\partial L_m}{\partial g^{\mu\nu}}.
\end{equation}

In consideration of the spatial isotropy and homogeneity of our Universe, we adopt the following flat Friedmann-Lema\^{i}tre-Robertson-Walker (FLRW) metric for our analysis,
\begin{equation}\label{9}
ds^2= -dt^2 + a^2(t)[dx^2+dy^2+dz^2],
\end{equation}
where $a(t)$ represents the scale factor that quantifies the cosmic expansion at a given time $t$. By considering the line element (\ref{9}), we have computed the Ricci scalar as
\begin{equation}\label{12}
R= 6 ( \dot{H}+2H^2 ),
\end{equation}
where $H=\frac{\dot{a}}{a}$ represents the Hubble parameter, $\dot{a}$ denotes the derivative of the scale factor $a$ with respect to time.

The energy-momentum tensor, which describes the matter content of the Universe modeled as a perfect fluid, exhibits the following non-zero components with respect to the line element \eqref{9},
\begin{equation}\label{13}
\mathcal{T}_{\mu\nu}=(\rho+p)u_\mu u_\nu + pg_{\mu\nu}.
\end{equation}

In the given context, the symbols $\rho$ represent the matter-energy density, $p$ denotes the spatially isotropic pressure, and $u^\mu=(1,0,0,0)$ represents the components of the four-velocity vector for the cosmic perfect fluid.

The Friedmann equations, which describe the dynamics of the Universe within the framework of $f(R,L_m)$ gravity, can be expressed as
\begin{equation}\label{14}
3H^2 f_R + \frac{1}{2} \left( f-f_R R-f_{L_m}L_m \right) + 3H \dot{f_R}= \frac{1}{2}f_{L_m} \rho,
\end{equation}
and
\begin{equation}\label{15}
\dot{H}f_R + 3H^2 f_R - \ddot{f_R} -3H\dot{f_R} + \frac{1}{2} \left( f_{L_m}L_m - f \right) = \frac{1}{2} f_{L_m}p.
\end{equation}  

\section{The $f(R,L_m)$ Cosmological Model }\label{sec3}

For our analysis, we consider the following functional form within the framework of the $f(R,L_m)$ cosmological model \cite{LB},
\begin{equation}\label{16} 
f(R,L_m)=\frac{R}{2}+L_m^\alpha,
\end{equation}
where $\alpha$ is the free parameter of the model. In particular, when $\alpha=1$, the standard Friedmann equations of GR are recovered. In the specific case of this $f(R,L_m)$ model, where $L_m=\rho$ \cite{HLR}, the Friedmann equations \eqref{14} and \eqref{15} can be expressed as 
\begin{equation}\label{17}
3H^2=(2\alpha-1) \rho^{\alpha},
\end{equation}
and
\begin{equation}\label{18}
2\dot{H}+3H^2=\left[(\alpha-1)\rho-\alpha p\right]\rho^{\alpha-1}.
\end{equation}

The effective EoS parameter ($w_{eff}$) is defined in terms of the total energy density ($\rho$) and pressure ($p$) as
\begin{equation}
    \omega_{eff}=\frac{p}{\rho}.
\end{equation}

The quantities $\rho$ and $p$ account for the density and pressure, respectively, encompassing all forms of matter present in the Universe. By using Eqs. (\ref{17}) and (\ref{18}), the effective EoS parameter can be expressed as
\begin{equation}
    \omega_{eff}=-1+\frac{(2-4 \alpha ) \dot{H}}{3 \alpha  H^2}.
    \label{eff}
\end{equation}

In order to find a solution for Eq. (\ref{eff}) and determine the value of $H$, an additional equation is required. To tackle this, we employ a carefully chosen parametric form for the equation of state parameter, which is expressed as a function of redshift $z$ \cite{EoS1},  $\omega_{eff}=-\frac{1}{1+m(1+z)^n}$, where $m$ and $n$ represent two model parameters. The form of $\omega_{eff}$ can be viewed as a phenomenological approach, providing a convenient and flexible parameterization of the EoS. This allows for model-independent analyses and enables investigations of a wide range of cosmological scenarios. As $z$ becomes very large (in the early Universe), the term $(1+z)^n$ dominates the denominator, leading to $\omega_{eff} \approx -\frac{1}{m(1+z)^n}$. Depending on the values of $m$ and $n$, the effective EoS can take different values in this high redshift regime. At moderate redshifts, where $(1+z)^n$ is still significant, but not dominant, the effective EoS is given by $\omega_{eff} = -\frac{1}{1+m(1+z)^n}$. This phase of the Universe is typically characterized by matter dominance, and $\omega_{eff}$ will be close to zero, representing non-relativistic matter behavior. As $z$ approaches zero (current epoch), the term $(1+z)^n$ becomes negligible, and $\omega_{eff} \approx -\frac{1}{1+m}$. In this phase, if $m>0$, the effective EoS will be negative, indicating accelerated expansion and behavior characteristic of DE domination. The specific values and behavior of $\omega_{eff}$ at different phases of the Universe are determined by the parameters $m$ and $n$. In the forthcoming section, we will aim to constrain these parameters using the most recent observational data available. By comparing the theoretical predictions based on the $\omega_{eff}$ model with the observational constraints, we can determine the range of values that are consistent with the data. Finally, it is important to note that this form of the equation of state parameter, $\omega_{eff}$, has been widely employed in numerous studies exploring various modified theories of gravity \cite{EoS2,EoS3,EoS4}.

From Eq. \eqref{eff}, we have
\begin{equation}\label{19}
\dot{H}+\frac{3 \alpha  H^2}{2 (2 \alpha -1)}-\frac{3 \alpha  H^2}{2(2 \alpha -1)(1+m (1+z)^n)}=0.
\end{equation}

By using the relation $ \frac{1}{H} \frac{d}{dt}= \frac{d}{dln(a)}$, we can express the given equation as a first-order differential equation,
\begin{equation}\label{20}
\frac{dH}{dln(a)}+\frac{3 \alpha  H}{2 (2 \alpha -1)}=\frac{3 \alpha  H}{2(2 \alpha -1)(1+m (1+z)^n)}.
\end{equation}

By integrating the aforementioned equation,  we can obtain the expression for the Hubble parameter in terms of redshift as
\begin{equation}\label{21}
H(z)=H_0 \left[A+B (1+z)^n\right]^l,
\end{equation}
where $H_0$ represents the current value of the Hubble parameter. It quantifies the rate of expansion of the Universe at the present cosmic epoch. In addition, $A=\frac{1}{1+m}$, $B=\frac{m}{1+m}$, and $l=\frac{3 \alpha }{2 n(2 \alpha-1)}$. It is worth mentioning that in the standard $\Lambda$CDM model, the parameters $\alpha=1$ and $n=3$ correspond to the specific case, with the present matter density parameter $\Omega_m=B=(1-A)$. As such, the values of the model parameters $\alpha$ and $n$ serve as indicators of the deviation of the current model from the $\Lambda$CDM model. These parameters allow us to quantify and compare the differences between the present model and the well-established $\Lambda$CDM framework, providing valuable insights into any potential modifications or extensions to the standard cosmological paradigm

The deceleration parameter is a fundamental parameter that plays a crucial role in describing the dynamics of the expansion phase of the Universe. It is defined as the negative ratio of the second derivative of the scale factor to its first derivative with respect to cosmic time,
\begin{equation}\label{22}
q=-1-\frac{\dot{H}}{H^2}.
\end{equation}

By substituting Eq. \eqref{21} into Eq. \eqref{22}, we obtain the following expression,
\begin{equation}\label{23}
q(z)=-1+\frac{B l n (1+z)^n}{A+B (1+z)^n}.
\end{equation}

\section{Constraints from Observational Data}\label{sec4}

In this section, we conduct a Bayesian analysis of the observational aspects of our cosmological model. We utilize two datasets, namely the CC dataset, and the Pantheon dataset, to determine the optimal ranges for the model parameters $H_0$, $m$, $n$, and $\alpha$. To constrain these parameters, we employ the standard Bayesian technique, utilizing the likelihood function and the Markov Chain Monte Carlo (MCMC) method implemented in the \texttt{emcee} Python library \cite{Mackey/2013}. The probability function used in this analysis aims to maximize the best-fit ranges of the model parameters, $\mathcal{L} \propto exp(-\chi^2/2)$, where the chi-squared function $\chi^2$ is employed to assess the goodness-of-fit between the model predictions and the observed data \cite{BS}. The specific form of the $\chi^2$ function varies depending on the dataset under consideration. Below, we provide the expressions for the $\chi^2$ functions used for different datasets.

\subsection{CC dataset}
In our analysis, we utilize Hubble parameter measurements obtained through the differential age method in the redshift range $0.07<z<2.42$ \cite{D1,D2,D3,D4}, commonly referred to as CC data. Specifically, we consider a dataset consisting of 31 data points compiled in \cite{Moresco/2015,Moresco/2018}. These measurements provide valuable information about the expansion history of the Universe and serve as important constraints for our cosmological model. In order to determine the best-fit values of the model parameters $H_0$, $m$, $n$, and $\alpha$, we define the chi-square function as
\begin{equation}\label{25}
\chi _{CC}^{2}(H_0,m,n,\alpha)=\sum\limits_{k=1}^{31}
\frac{[H_{th}(z_{k},H_0,m,n,\alpha)-H_{obs}(z_{k})]^{2}}{
\sigma _{H(z_{k})}^{2}},  
\end{equation}
where, $H_{th}$ corresponds to the theoretical value of the Hubble parameter derived from our model, while $H_{obs}$ represents its observed value. The standard deviation, denoted as $\sigma_{H(z_{k})}$, quantifies the uncertainty associated with the Hubble parameter measurements. Fig. \ref{F_CC} illustrates the likelihood contours for the model parameters $H_0$, $m$, $n$, and $\alpha$ based on the CC dataset, specifically showing the $1-\sigma$ and $2-\sigma$ contours. The best-fit values obtained for the model parameters are $H_0=67.86_{-0.75}^{+0.75}$, $m=0.48_{-0.14}^{+0.14}$, $n=5.8_{-3.7}^{+3.8}$, and $\alpha=1.33_{-0.45}^{+0.45}$.

\subsection{Pantheon dataset}

In recent studies, a comprehensive compilation of Pantheon SNe Ia data has been made available. The Pantheon dataset comprises 1048 data points, incorporating observations from various surveys such as PanSTARSS1 Medium, SDSS, SNLS, and Deep Survey, as well as low redshift and HST surveys. Scolnic et al. \cite{Scolnic/2018} have meticulously assembled this extensive dataset covering a wide redshift range of $z \in [0.01,2.3]$. Considering a spatially flat Universe based on the findings of the Planck collaboration \cite{planck_collaboration/2020}, the luminosity distance can be expressed as 
\begin{equation}\label{26}
D_{L}(z)= c(1+z) \int_{0}^{z} \frac{ dz'}{H(z')},
\end{equation}
where $c$ is the speed of light.

To perform statistical analysis, we employ the $\chi^{2}$ function to assess the agreement between the observed supernovae samples and the theoretical predictions. The $\chi^{2}$ function is expressed as
\begin{equation}\label{27}
\chi^2_{Pantheon}(H_0,m,n,\alpha)=\sum_{i,j=1}^{1048}\Delta\mu_{i}\left(C^{-1}_{Pantheon}\right)_{ij}\Delta\mu_{j},
\end{equation}
where $C_{Pantheon}$ represents the covariance metric \cite{Scolnic/2018}, and
\begin{equation}
  \quad \Delta\mu_{i}=\mu^{th}(z_i,H_0,m,n,\alpha)-\mu_i^{obs}.
\end{equation}

Here, the symbol $\mu^{th}$ represents the theoretical value of the distance modulus, while $\mu^{obs}$ corresponds to its observed value. The distance modulus is calculated theoretically as
\begin{equation}\label{28}
\mu^{th}(z)= 5log_{10}D_{L}(z)+\mu_{0}, 
\end{equation}
with 
\begin{equation}\label{29}
\mu_{0} =  5log(1/H_{0}Mpc) + 25,
\end{equation}

Fig. \ref{F_SN} illustrates the likelihood contours for the model parameters $H_0$, $m$, $n$, and $\alpha$ based on the Pantheon data sample, specifically showing the $1-\sigma$ and $2-\sigma$ contours. The best-fit values obtained for the model parameters are $H_0=67.89_{-0.78}^{+0.80}$, $m=0.48_{-0.14}^{+0.14}$, $n=5.4_{-3.8}^{+4.0}$, and $\alpha=1.33_{-0.47}^{+0.50}$.

\begin{widetext}

\begin{figure}[h]
\centerline{\includegraphics[scale=0.70]{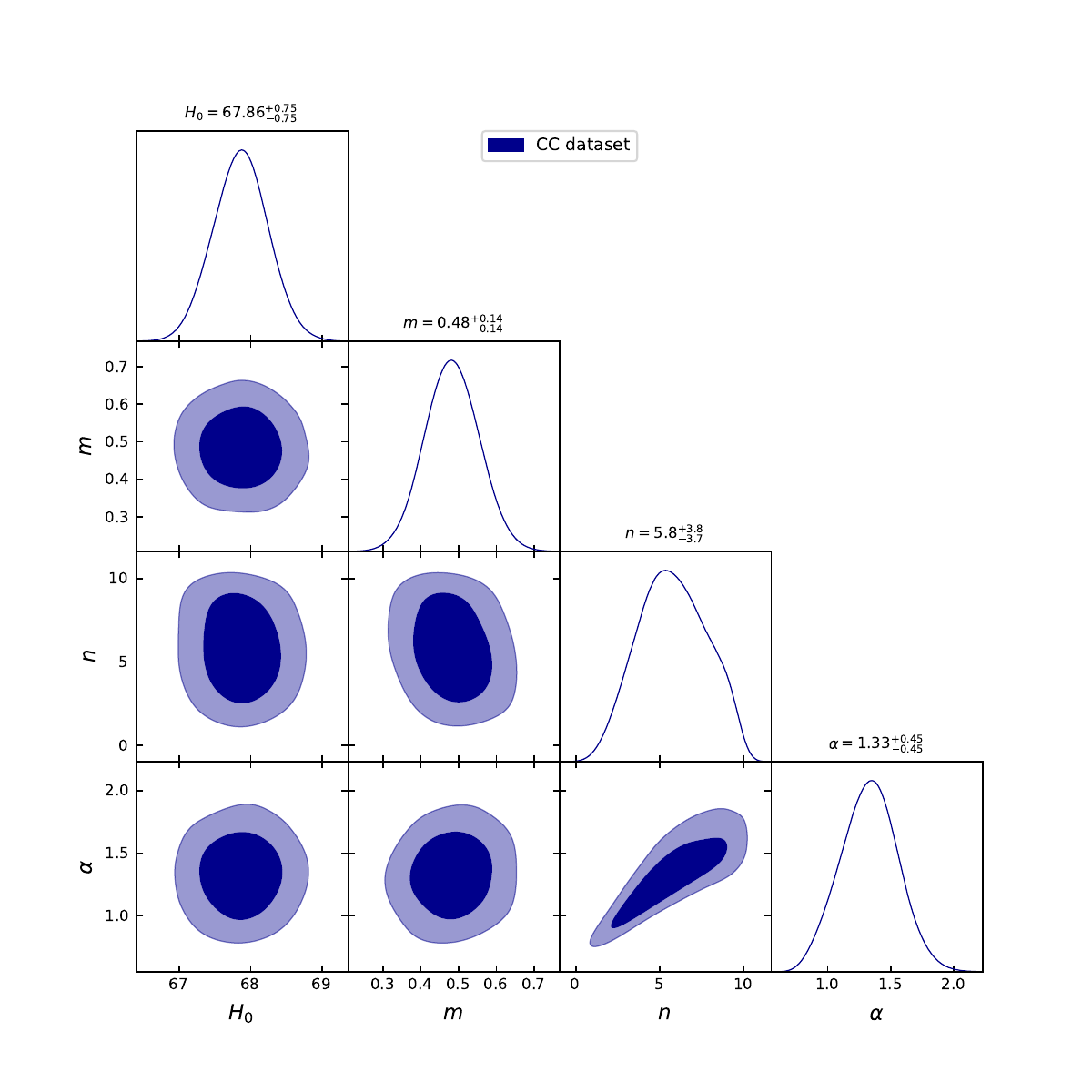}}
\caption{Likelihood contours for model parameters using CC dataset: $1-\sigma$ and $2-\sigma$ confidence intervals.}
\label{F_CC}
\end{figure}

\begin{figure}[h]
\centerline{\includegraphics[scale=0.70]{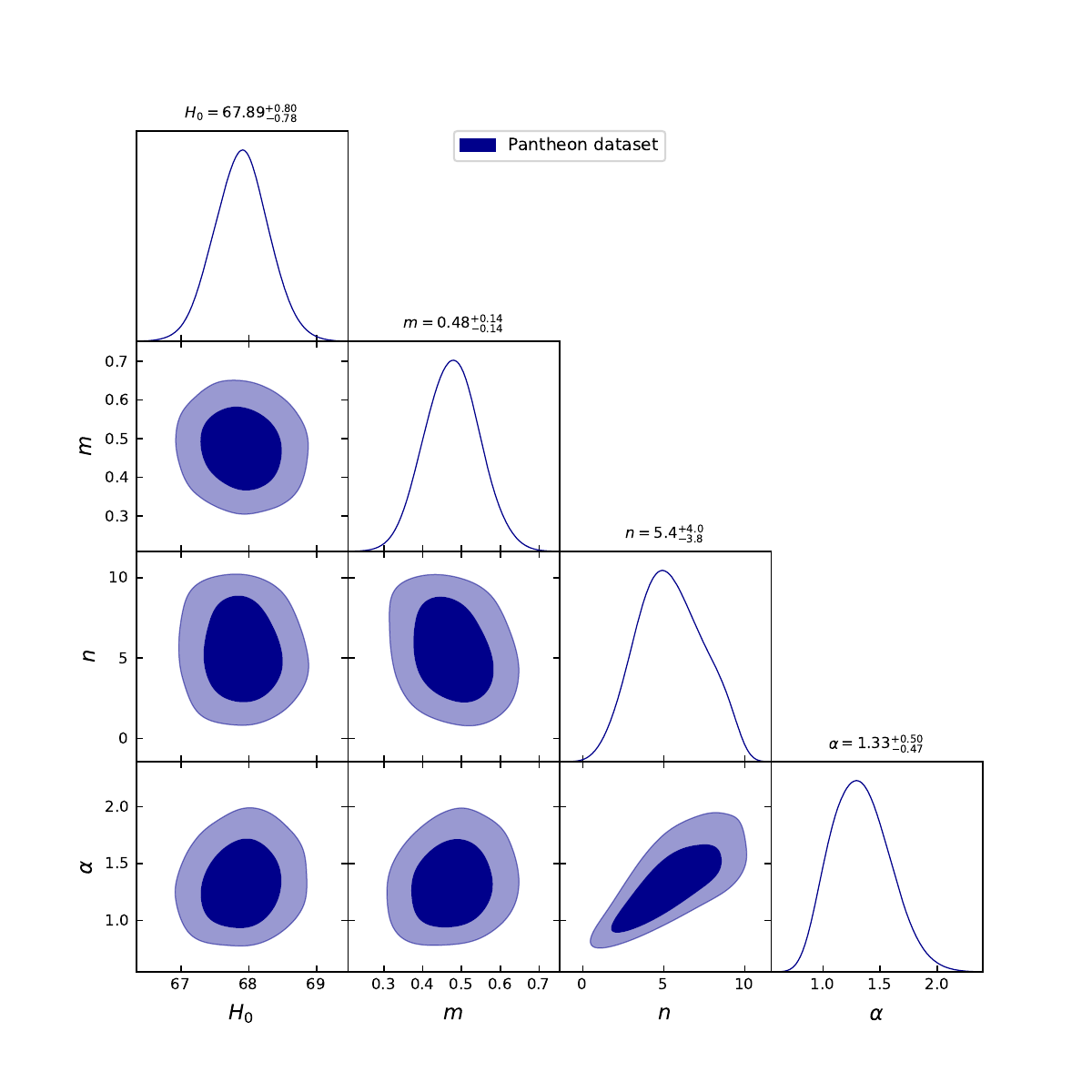}}
\caption{Likelihood contours for model parameters using Pantheon data sample: $1-\sigma$ and $2-\sigma$ confidence intervals.}
\label{F_SN}
\end{figure}
\end{widetext}

The evolution of the effective EoS, density, pressure, and deceleration parameters is shown below, considering the constrained values of the model parameters.

\begin{figure}[H]
\includegraphics[scale=0.7]{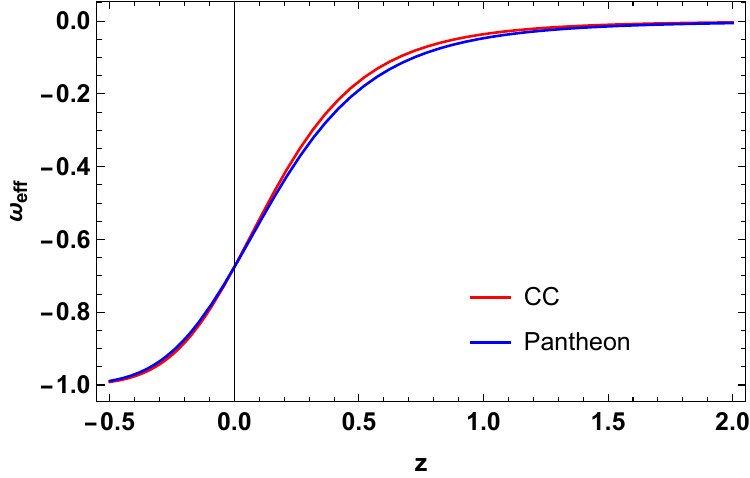}
\caption{Evolution of the effective EoS parameter with cosmic redshift for constrained model parameters using CC and Pantheon datasets.}\label{F_EoS}
\end{figure}

The EoS parameter plays a crucial role in determining the relationship between energy density and pressure in different phases of the Universe. Some common phases observed through the EoS parameter include the dust phase ($\omega = 0$), the radiation dominated phase ($\omega = 1/3$), and the vacuum energy phase ($\omega = -1$) corresponding to the $\Lambda$CDM model. In addition, there is the accelerating phase of the Universe, which is a topic of recent discussion, and is characterized by $\omega < -1/3$. This phase includes the quintessence regime ($-1 < \omega \leq -1/3$) and the phantom regime ($\omega < -1$). In this study, we consider an effective EoS parameter that depends on two model parameters, $m$ and $n$. Based on the constrained values of these parameters from the CC and Pantheon datasets, the behavior of the effective EoS parameter is shown in Fig. \ref{F_EoS}. At $z = 0$, the value of the effective EoS parameter is determined to be $\omega_{eff}^{0}=-0.68\pm0.06$ for both the CC and Pantheon datasets \cite{EoS4,Gruber}, indicating a quintessence phase. This value is obtained through the analysis of observational data and represents the behavior of the cosmic expansion at the present epoch. The quintessence phase is characterized by $-1 < \omega \leq -1/3$, indicating the presence of a form of DE that drives the accelerated expansion of the Universe. 
It is important to note that both datasets exhibit similar behavior, as indicated by the consistent values of the effective EoS parameter obtained from the analysis.

\begin{figure}[H]
\includegraphics[scale=0.7]{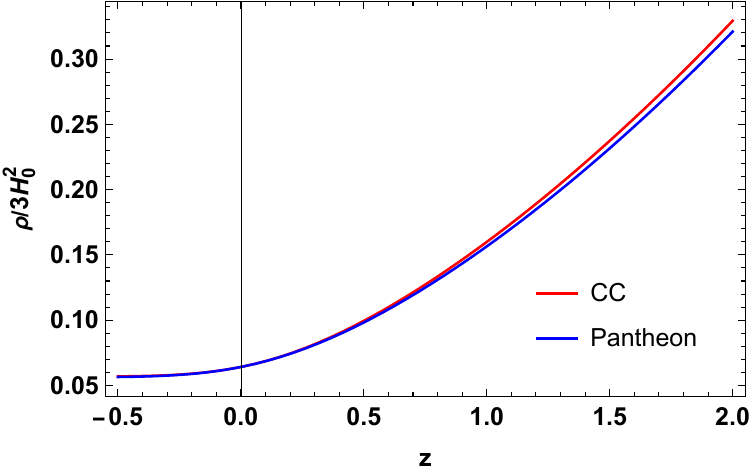}
\caption{Evolution of the energy density parameter with cosmic redshift for constrained model parameters using CC and Pantheon datasets.}\label{F_rho}
\end{figure}

\begin{figure}[H]
\includegraphics[scale=0.7]{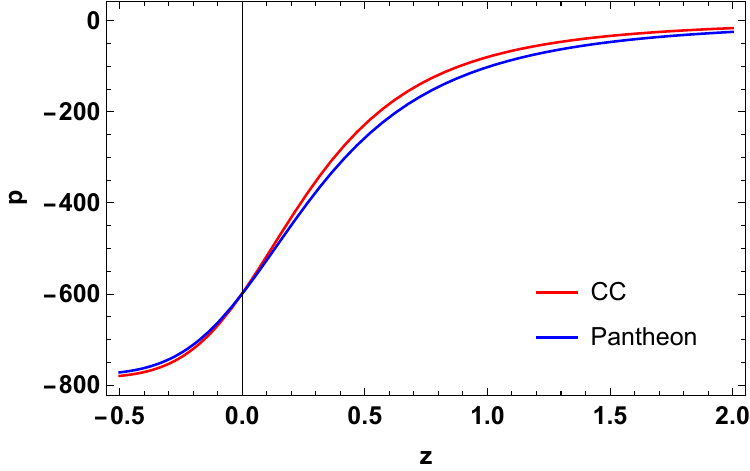}
\caption{Evolution of the pressure with cosmic redshift for constrained model parameters using CC and Pantheon datasets.}\label{F_p}
\end{figure}

Figs. \ref{F_rho} and \ref{F_p} illustrate the evolution of the energy density and pressure, respectively. It is noteworthy that the energy density exhibits the expected positive behavior, indicating a contribution to the Universe's expansion. On the other hand, the pressure displays a negative behavior both in the present and future, suggesting a driving force for the expansion. This observation suggests that the effective EoS parameter, augmented by the additional terms in the $f(R,L_m)$ gravity, contributes to the accelerated expansion of the Universe.

\begin{figure}[H]
\includegraphics[scale=0.7]{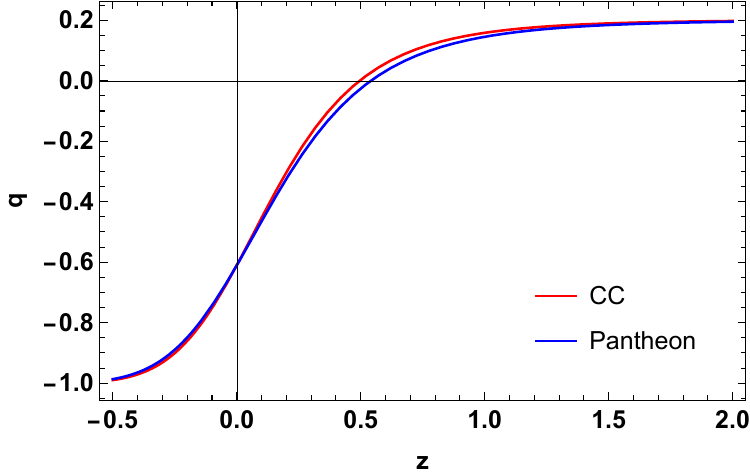}
\caption{Evolution of the deceleration parameter with cosmic redshift for constrained model parameters using CC and Pantheon datasets.}\label{F_q}
\end{figure}

In Fig. \ref{F_q}, we observe that the Universe follows a specific evolution pattern. It starts its history in a decelerating phase ($q > 0$), indicating a slowing down of the expansion. However, after a transition redshift $z_tr$ (i.e. at $q=0$), the Universe enters an accelerating phase ($q < 0$), characterized by an increasing rate of expansion. This behavior aligns with the observed dynamics of the recent Universe, which has gone through distinct stages including a decelerating dominated phase, an accelerating expansion phase, and a late-time accelerating phase. It is important to note that the evolution of the Universe eventually leads to a de Sitter expansion, characterized by a constant and asymptotically lower value of the deceleration parameter at lower redshifts (i.e. at $z \to -1$).

The transition redshifts, determined by the constrained values of the model parameters using the CC and Pantheon datasets, are found to be $z_t = 0.50^{+0.04}_{-0.04}$ and $z_t = 0.54^{+0.02}_{-0.04}$, respectively \cite{Jesus,Garza}. This indicates a shift in the dynamics of the Universe from a decelerating phase to an accelerating phase at this particular redshift. 
Furthermore, the present value of the deceleration parameter is estimated to be $q_0 = -0.61^{+0.01}_{-0.01}$ for both the CC and Pantheon datasets \cite{Lu,Campo}. This negative value suggests that the Universe is currently experiencing an accelerating expansion.

\section{Statefinder parameters}\label{sec5}

To analyze and characterize cosmic acceleration, various models of dark energy (DE) have been developed. To distinguish between these different cosmological scenarios involving DE, it is important to have a sensitive and reliable diagnostic tool. One such diagnostic approach is the "statefinder" introduced by Sahni et al. \cite{V.S.}. The statefinder is designed to probe the dynamics of the Universe's expansion by utilizing higher derivatives of the scale factor, building upon the information provided by the Hubble parameter and the deceleration parameter. The statefinder diagnostic, represented by the parameter pair $(r, s)$, offers a valuable tool for distinguishing between different DE models. Various cosmological models, including DE, exhibit distinct evolutionary paths in the $r-s$ plane. For instance, the spatially flat $\Lambda$CDM model corresponds to $(r=1, s=0)$. The quintessence model is characterized by $(r<1, s>0)$, the Chaplygin gas model corresponds to $(r>1, s<0)$, and the holographic DE model is represented by $(r=1, s=\frac{2}{3})$.

The statefinder parameters, $(r, s)$, are defined as
\begin{equation}
r=\frac{\overset{...}{a}}{aH^{3}},
\end{equation}%
\begin{equation}
s=\frac{\left( r-1\right) }{3\left( q-\frac{1}{2}\right) }.
\end{equation}

The parameter $r$ can be rewritten as%
\begin{equation*}
r=2q^{2}+q-\frac{\overset{.}{q}}{H}.
\end{equation*}%

\begin{figure}[H]
\includegraphics[scale=0.7]{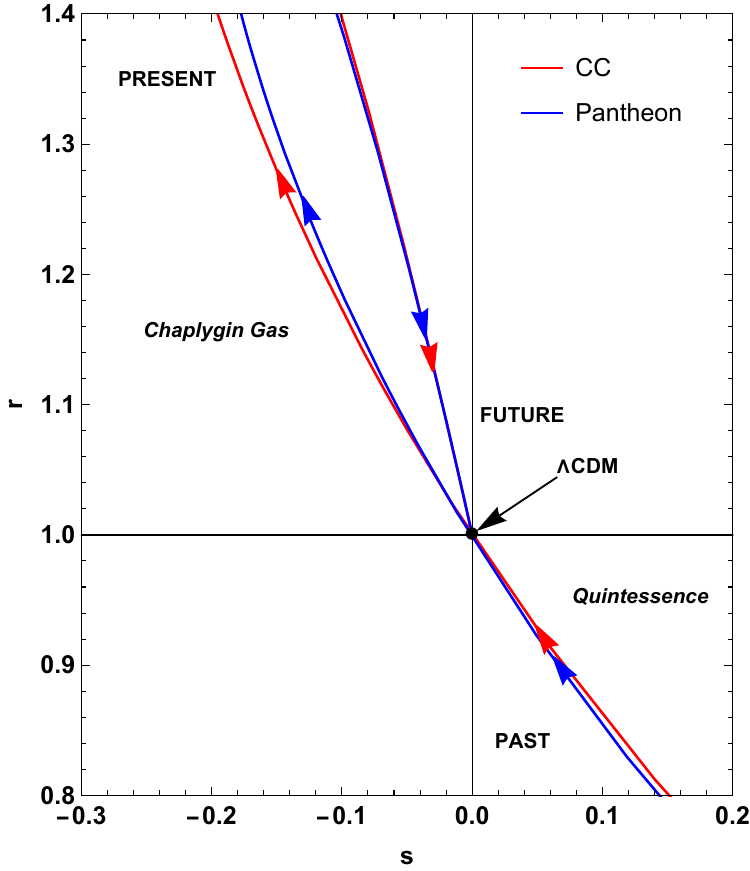}
\caption{Evolution of the $s-r$ plane for constrained model parameters using CC and Pantheon datasets.}\label{F_rs}
\end{figure}

\begin{figure}[H]
\includegraphics[scale=0.7]{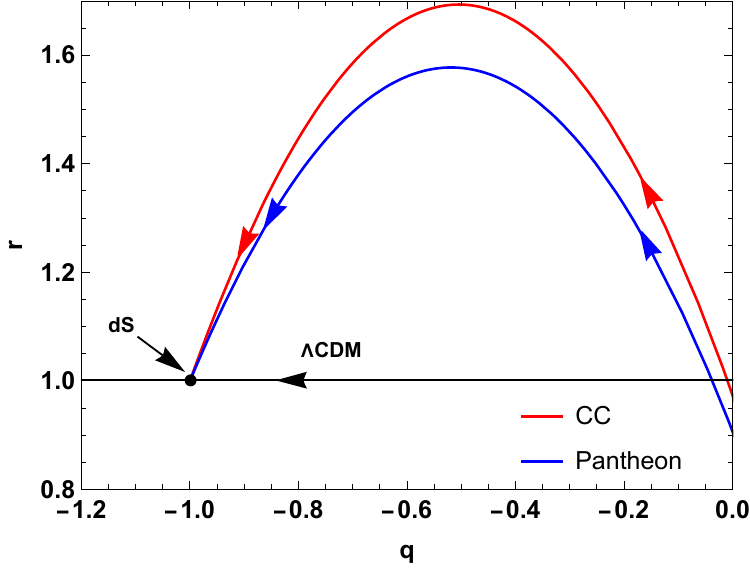}
\caption{Evolution of the $q-r$ plane for constrained model parameters using CC and Pantheon datasets.}\label{F_rq}
\end{figure}

In view of Figs. \ref{F_rs} and \ref{F_rq}, the evolution of the statefinder pairs $s-r$ and $q-r$ exhibits a trajectory that starts in the quintessence regime at early times, transitions through the Chaplygin gas region, and ultimately approaches the $\Lambda$CDM point. In the $s-r$ plane, the fixed point $(0,1)$ represents the spatially flat $\Lambda$CDM model, while in the $q-r$ plane, the point $(-1,1)$ corresponds to the de Sitter (dS) point. This behavior demonstrates that the statefinder diagnostic is a more effective quantity than the EoS for distinguishing between different DE models in this scenario.

\section{Conclusion}\label{sec6}

Although numerous observations have confirmed the existence of DE, its underlying nature still eludes us. The EoS parameter plays a crucial role in characterizing various DE models. It serves as a significant method for understanding and distinguishing the nature of DE and its effects on the expansion of the Universe. The condition for accelerating expansion is characterized by an EoS parameter $\omega < -1/3$. Understanding the gravitational and dynamic aspects of the Universe hinges on unraveling the fundamental physics behind DE, which directly influences EoS.

In this study, we have pursued a novel approach to the gravitational theory known as $f(R, L_m)$ gravity. This alternative framework offers a promising avenue for exploring modified theories of gravity. To analyze the cosmological implications of this theory, we have adopted a parametric form for the EoS parameter that depends on the redshift $z$, $\omega_{eff}=-\frac{1}{1+m(1+z)^n}$. This choice allows us to investigate the behavior of EoS and its impact on the evolution of the Universe within the context of $f(R, L_m)$ gravity. 
At the epoch of recent acceleration, the effective EoS parameter in our model takes a negative value, indicating a driving force behind the accelerated expansion of the Universe. Specifically, $\omega_{eff}$ is less than $-1/3$, which is a characteristic condition for cosmic acceleration. As we move towards higher redshifts, the value of $\omega_{eff}$ tends to approach zero. This behavior is observed when the model parameters $m$ and $n$ are positive, and the specific value of $\omega_{eff}$ at $z = 0$ is determined by these model parameters. 
Using the parametric form mentioned above, we have obtained the solution for the Hubble parameter in a non-linear $f(R,L_m)$ model. In this particular model, the functional form is given by $f(R,L_m)=\frac{R}{2}+L_m^\alpha$, where $\alpha$ represents a free model parameter.

In addition, we used two datasets, namely the CC dataset with 31 data points and the Pantheon dataset with 1048 data points, to determine the optimal values for the model parameters $(H_0, m, n, \alpha)$, which are found to be: $H_0=67.86_{-0.75}^{+0.75}$, $m=0.48_{-0.14}^{+0.14}$, $n=5.8_{-3.7}^{+3.8}$, $\alpha=1.33_{-0.45}^{+0.45}$ for the CC dataset and $H_0=67.89_{-0.78}^{+0.80}$, $m=0.48_{-0.14}^{+0.14}$, $n=5.4_{-3.8}^{+4.0}$, $\alpha=1.33_{-0.47}^{+0.50}$ for the Pantheon dataset. Furthermore, we have examined the dynamics of the effective EoS, density, pressure, and deceleration parameters, taking into account the constrained values of the model parameters. Based on the analysis presented in Fig. \ref{F_EoS}, the effective EoS parameter at present ($z = 0$) is found to be $\omega_{eff}^{0}=-0.68\pm0.06$ for both the CC and Pantheon datasets. This result suggests that the Universe is currently in a quintessence phase. Figs. \ref{F_rho} and \ref{F_p} depict the evolution of energy density and pressure. The energy density exhibits a positive behaviour, whereas the pressure displays a negative behavior, acting as a driving force for the expansion. The deceleration parameter in Fig. \ref{F_q} shows a recent transition of the Universe from deceleration to acceleration. The transition redshifts determined by the constrained model parameters are $z_t = 0.50^{+0.04}_{-0.04}$ for the CC dataset and $z_t = 0.54^{+0.02}_{-0.04}$ for the Pantheon dataset. Moreover, the present values of the deceleration parameter are $q_0 = -0.61^{+0.01}_{-0.01}$ for both data sets.

Furthermore, we investigated the statefinder parameters for our model, taking into account observational constraints. The results, shown in Figs. \ref{F_rs} and \ref{F_rq}, indicate that our cosmological $f(R,L_m)$ model exhibits a trajectory starting in the quintessence regime at early times, transitioning through the Chaplygin gas region, and eventually approaching the $\Lambda$CDM point. This behavior highlights the effectiveness of the statefinder diagnostic in distinguishing between different DE models in this particular scenario. In this study, our primary focus has been on the non-linear $f(R,L_m)$ model, specifically the minimal coupling case represented by $f(R,L_m)=\frac{R}{2}+L_m^\alpha$. This particular model has provided valuable insights into understanding the dynamics of the Universe within the context of modified gravity theories. However, it's crucial to acknowledge that our exploration is not exhaustive, and there are numerous uncharted models within the realm of $f(R,L_m)$ gravity that merit investigation. One intriguing example is the non-minimal coupling case, characterized by the functional form $f(R,L_m)=\frac{R}{2}+(1+\alpha R)L_m$. This alternative model, as discussed in a previous study \cite{RL2}, offers a unique perspective on how the Universe's evolution is influenced by non-linear $f(R,L_m)$ gravity. Future research endeavors can extend their inquiries to this specific model, delving into its complexities and implications. By doing so, we can gain a more comprehensive understanding of cosmic dynamics, considering various facets of matter-geometry interactions and their role in shaping the Universe's behavior. Exploring different models within the framework of $f(R,L_m)$ gravity is a promising avenue for advancing our knowledge of the Universe's fundamental processes.

\section*{Data Availability Statement}
There are no new data associated with this article.

\section*{Acknowledgments} 
This work was supported and funded by the Deanship of Scientific Research at Imam Mohammad Ibn Saud Islamic University (IMSIU) (grant number IMSIU-RP23007).


\end{document}